\pdfoutput=1

\documentclass[11pt]{article}

\usepackage[preprint]{acl}

\usepackage{times}
\usepackage{latexsym}

\usepackage[T1]{fontenc}

\usepackage[utf8]{inputenc}

\usepackage{microtype}

\usepackage{inconsolata}

\usepackage{graphicx}
\usepackage{enumitem}
\usepackage{multirow}

\usepackage{times}
\usepackage{latexsym}
\usepackage{textcomp}
\usepackage[T1]{fontenc}
\usepackage[utf8]{inputenc}
\usepackage{microtype}
\usepackage{inconsolata}
\usepackage{booktabs}
\usepackage{xspace}
\usepackage{amsmath}
\usepackage{amssymb}
\usepackage{mathtools}
\usepackage{tcolorbox}
\usepackage{xspace}
\usepackage{caption}
\usepackage{subcaption}
\usepackage{balance}
\usepackage[hang,flushmargin]{footmisc}
\usepackage{multirow}

\usepackage[hang, flushmargin]{footmisc}

\title{
Rank-R1: Enhancing Reasoning in LLM-based Document Rerankers via Reinforcement Learning
}

 \author{Shengyao Zhuang$^{*,1}$, Xueguang Ma$^{*,2}$, Bevan Koopman$^{1,3}$, Jimmy Lin$^{2}$, Guido Zuccon$^{3}$ \\
         $^1$CSIRO, \\ $^2$ University of Waterloo,
         \\ $^3$ The University of Queensland}
     
\begin{document}
\maketitle
\def\thefootnote{*}\footnotetext{These authors contributed equally to this work. Work in progress.}\def\thefootnote{\arabic{footnote}}

\begin{abstract}

In this paper, we introduce Rank-R1, a novel LLM-based reranker that performs reasoning over both the user query and candidate documents before performing the ranking task.

Existing document reranking methods based on large language models (LLMs) typically rely on prompting or fine-tuning LLMs to order or label candidate documents according to their relevance to a query. For Rank-R1, we use a reinforcement learning algorithm along with only a small set of relevance labels (without any reasoning supervision) to enhance the reasoning ability of LLM-based rerankers. Our hypothesis is that adding reasoning capabilities to the rerankers can improve their relevance assessement and ranking capabilities.  

Our experiments on the TREC DL and BRIGHT datasets show that Rank-R1 is highly effective, especially for complex queries. In particular, we find that Rank-R1 achieves effectiveness on in-domain datasets at par with that of supervised fine-tuning methods, but utilizing only 18\% of the training data used by the fine-tuning methods.
We also find that the model largely outperforms zero-shot and supervised fine-tuning when applied to out-of-domain datasets featuring complex queries, especially when a 14B-size model is used. Finally, we qualitatively observe that Rank-R1's reasoning process improves the explainability of the ranking results, opening new opportunities for search engine results presentation and fruition.

\end{abstract}

\section{Introduction}

\noindent Large language models (LLMs) have shown strong performance in document ranking tasks~\cite{ma2023zeroshotlistwisedocumentreranking,sun-etal-2023-chatgpt,zhuang-etal-2023-open,setwise,zhuang-etal-2024-beyond,sun2024investigation,xu2024rankmambabenchmarkingmambasdocument}. Zero-shot prompting methods such as RankGPT, rerank documents by directly prompting LLMs to generate reordered document lists~\cite{sun-etal-2023-chatgpt}. However, these methods rely on the assumption that the LLM follows instructions well.
Moreover, being zero-shot, they do not leverage available human-annotated relevance data for further improvement.

In contrast, methods like RankLlama fine-tune LLMs using human relevance judgments, assigning scores to each query-document pair~\cite{rankllama}.
While effective, these approaches do not explicitly model reasoning processes.
This is mainly due to the lack of high-quality reasoning data for supervised fine-tuning.
In practice, user relevance judgments often come in the form of selecting the most relevant document from a set, but achieving high-quality rankings requires strong reasoning capabilities to interpret complex relevance relationships.

Recent advancements in reinforcement learning (RL) for LLMs, such as DeepSeek-R1~\cite{deepseekr1} and Simple RL~\cite{zeng2025simplerl}, have demonstrated that reward-based training can enhance reasoning abilities, particularly in tasks like mathematical question-answering~\cite{deepseekmath}. These recent innovations suggest that rule-based reward scoring alone can improve an LLM's ability to reason and explain.

Building on this insight, we ask whether reinforcement learning in the style of Deepseek-R1 can improve reasoning in document reranking.
Specifically, we apply Group Relative Policy Optimization (GRPO)~\cite{deepseekmath} to train an LLM-based reranker.
Given a user query and a list of retrieved candidate documents, the reranker generates reasoning steps before selecting the most relevant document.
The reward signal is determined only by whether the model eventually identifies the most relevant document among the candidates correctly.

We train our model, Rank-R1, on the MS MARCO passage ranking dataset and evaluate it on TREC DL19 and DL20 (in-domain datasets).
Our results show that RL-based training is at par with supervised fine-tuning on in-domain data. 

Additionally, we evaluate Rank-R1 on the BRIGHT dataset~\cite{su2025bright}, which requires complex query reasoning and relevance understanding, and is out-of-domain with respect to the data used to train the model.
Rank-R1, trained with retrieval reinforcement learning, outperforms both zero-shot prompting and supervised fine-tuning on this out-of-domain dataset.
Notably, our 14B model surpasses the much larger (zeroshot) GPT-4 in reranking performance on the BRIGHT dataset.
\section{Method}
To train Rank-R1, we adapt the RL training framework proposed by DeepSeek~\cite{deepseekmath,deepseekr1} to enhance the reasoning ability of LLM-based document rerankers. In this section, we discuss the details of each component in our method.

\subsection{LLM Reranking} 

Our LLM reranker follows the Setwise prompting approach proposed by \citet{setwise}. This method takes a query and a set of candidate documents as input to the LLM and prompts the LLM to select the most relevant document among the candidates based on relevance to the query. Then, the heapsort algorithm is used to build a heap tree over all the candidate documents from the first-stage retriever, and the documents are reranked via the ``heapify'' operations with the Setwise prompt.

However, the original Setwise ranking approach does not encourage the LLM to reason about the relevance between the query and the documents. Instead, it directly asks for the most relevant one. To unlock the reranker's reasoning ability, we modify the original Setwise prompt by adding a reasoning instruction, as shown in Figure~\ref{fig:prompt}.
Specifically, we adapt the system prompt from the DeepSeek-R1-Zero method to the Setwise prompt. This modification encourages LLMs to reason first before providing an answer—in our case, predicting the label of the most relevant candidate document to the query. We refer to the Setwise method using this modified prompt as \textit{Rank-R1}.

\begin{figure}[h]
\begin{tcolorbox}
\small
SYSTEM:\\
A conversation between User and Assistant. The user asks a question, and the Assistant solves it. The assistant first thinks about the reasoning process in the mind and then provides the user with the answer. The reasoning process and answer are enclosed within <think> </think> and <answer> </answer> tags, respectively, i.e., <think> reasoning process here </think> <answer> answer here </answer>.\\

USER:\\
Given the query: "\{query\}", which of the following documents is most relevant?\\
\textnormal{[1]} \{document1\}\\
\textnormal{[2]} \{document2\}\\
....\\
\textnormal{[20]} \{document20\}

After completing the reasoning process, please provide only the label of the most relevant document to the query, enclosed in square brackets, within the answer tags. For example, if the third document is the most relevant, the answer should be: <think> reasoning process here </think> <answer>[3]</answer>.
\end{tcolorbox}
\caption{Prompt used for Rank-R1.}
\label{fig:prompt}
\end{figure}

\subsection{Reinforcement Learning}
Although any modern instruction-tuned LLM, when coupled with our \textit{Rank-R1} prompt, may exhibit strong zeroshot reasoning ability, their reasoning process could still be suboptimal for the Setwise ranking method. This is because the LLMs might not have been fine-tuned on similar instructional data. On the other hand, gathering human-annotated reasoning data for large-scale supervised fine-tuning of  \textit{Rank-R1} could be both costly and infeasible. To address these challenges, we employ the GRPO RL algorithm~\cite{deepseekmath} to enhance the reasoning process of the \textit{Rank-R1} reranker. The GRPO algorithm optimizes the following objective:

\begin{small}
\begin{eqnarray}
\small
    \mathcal{J}_{GRPO}(\theta) &=&\mathbb{E}[q \sim P(Q), \{o_i\}_{i=1}^{|G|} \sim \pi_{\theta_{old}}(O|q)] \nonumber\\
    && \frac{1}{|G|}\sum_{i=1}^{|G|} \bigg( \min \Big( \frac{\pi_\theta(o_i |q)}{\pi_{\theta_{old}}(o_i |q)} A_i,\nonumber\\
    &&\text{clip} \big( \frac{\pi_\theta(o_i |q)}{\pi_{\theta_{old}}(o_i |q)}, 1 - \epsilon, 1 + \epsilon \big)  A_i \Big) \nonumber\\
    && - \beta \mathbb{D}_{KL}(\pi_{\theta} || \pi_{ref})\bigg)
\label{eq:GRPO-obj}
\end{eqnarray}
\end{small}
\noindent where $Q$ is the Setwise ranking prompt in the training data, $G$ is a group of generated data points sampled from an old policy $\pi_{\theta_{old}}$ (in our case, the initial LLM) given a sampled prompt, and $\pi_\theta$ is the LLM ranker we are optimizing. The clip operation is used to cap the ratio of the new and old policies. In our experiment, we only use the data samples from $\pi_{\theta_{old}}$ to update $\pi_{\theta}$ once, meaning $\pi_{\theta_{old}} = \pi_{\theta}$, so Eq~\ref{eq:GRPO-obj} reduces to:

\begin{small}
\begin{align}
    &\mathcal{J}_{GRPO}(\theta) = \nonumber\\
    &\frac{1}{|G|}\sum_{i=1}^{|G|} \left( \frac{\pi_\theta(o_i |q)}{\pi_{\theta_{old}}(o_i |q)} A_i, - \beta \mathbb{D}_{KL}(\pi_{\theta} || \pi_{ref})\right)
    \label{eq:GRPO-obj-simple}
\end{align}
\end{small}
\noindent where $\mathbb{D}_{KL}$ is the KL loss that penalizes how far the new policy differs from a reference policy $\pi_{ref}$, which in our case is the original instruction-tuned LLM.

The generated data points, in our case, represent the reasoning process. We use the generated relevant document label to calculate the advantage $A$ as follows:
\begin{equation}
    A_i = \frac{r_i - {\mathrm mean(\{r_1, r_2, \cdots, r_G\})}}{{\mathrm std(\{r_1, r_2, \cdots, r_G\})}}
\end{equation}
where $r$ is the reward given by the training data, which we describe in detail later. Overall, the GRPO training optimizes the LLM to generate tokens that maximize the rewards.

The Setwise prompts and rewards for our GRPO training come from training data. For constructing such RL training data, we use the MSMARCO training data\footnote{\url{https://huggingface.co/datasets/Tevatron/msmarco-passage}} provided by the Tevatron IR toolkit~\cite{tevatron}, which includes training queries, human-labeled relevant documents, and BM25-retrieved top-100 documents. For each training query, we sample 19 documents from the set retrieved by BM25, along with one labeled relevant document, to form the \textit{Rank-R1} prompt.




We design the reward function for RL as follows: a reward of one is granted if and only if the LLM generations match the reasoning and answering format (i.e., the generated tokens fill in the <think> </think> <answer> </answer> spans) \textit{and} the answer correctly matches the label of the ground-truth relevant document. Otherwise, a reward of zero is provided. Our straightforward rule-based reward mechanism does not impose constraints on the reasoning process but instead encourages the model to generate the correct format and answer following the reasoning. Most importantly, this RL training does not require human-annotated reasoning data.

There are two main reasons why we selected the Setwise ranking approach as the backend of our Rank-R1: First, the nature of Setwise prompting allows us to use a simple rule-based reward function, as only the most relevant document label needs to be predicted, enabling a straightforward match with the answer span. Secondly, it is convenient for us to fairly compare the RL-trained Rank-R1 with the supervise fine-tuned Setwise reranker, which simply removes the reasoning process from the prompt and is directly trained to predict the ground-truth label using the same training data. Other LLM-based ranking methods, such as Listwise prompting, are harder to fairly compare in this setting, as a ground-truth ranking is usually not available in the training data (MSMARCO only has one judged relevant document per query on average).

\section{Experimental Settings}

\paragraph{Datasets.} We explore to dataset settings to evaluate the effectiveness of Rank-R1: an in-domain setting, where we use the TREC-DL19 and DL20 datasets~\cite{dl19,dl20}, and an out-of-domain setting, where we use the BRIGHT benchmark datasets~\cite{su2025bright}. The DL19 and DL20 are in-domain because they are based on the MSMARCO passage ranking dataset -- the same dataset used in the training of our rerankers; unlike MSMARCO though they contain deep assessments for each query (around 210 assessments per query on average).  
The BRIGHT benchmark datasets encompass domains, including biology, code, and math, and require intensive reasoning to rank relevant documents.

\paragraph{Initial Retrieval.} For all methods we consider in our experiments, the initial retrieval is performed using the \textit{pyserini} implementation of BM25~\cite{pyserini}. Reranking approaches are then provided the top 100 documents retrieved by BM25 to rerank.

\paragraph{Rank-R1 Settings.} We compare two settings of Rank-R1: (i) a Zeroshot setting, where only the Setwise-based prompt, improved by the presence of the reasoning instruction is used, with no training performed, and (ii) a GRPO setting, where we employ the same prompt as in the Zeroshot setting, but we also train the model according to the GRPO RL method. For GRPO, we set $|G| = 8$, that is, for each training query we generate 8 answers with the old policy $\pi_{\theta_{old}}$; as training dataset we use the training part of MS MARCO. 
For all Rank-R1 experiments, due to the limited computational resources, we only train on approximately 18\% of the full MSMARCO dataset, requiring roughly three (3B and 7B models) to five (14B model) days of training on four H100 GPUs. The details of GRPO training hyperparameters are provided in Appendix~\ref{sec:appendix_train}.

\paragraph{Comparison Methods.} To understand the effect of the reasoning prompt and the GRPO training on the Setwise approach, we compare Rank-R1 against the original Setwise method (using prompt as listed in Appendix~\ref{sec:appendix}), either used in a Zeroshot manner, or trained with the standard supervised fine-tuning (SFT) on MS MARCO data (400k training datapoints circa). The details of SFT training hyperparameters are provided in Appendix~\ref{sec:appendix_train}.

To further contextualise the effectiveness of Rank-R1, we also consider the effectiveness of the current state-of-the-art Listwise reranker, RankZephyr~\cite{pradeep2023rankzephyreffectiverobustzeroshot}, which was trained with data generated from GPT-4. We use the model checkpoint\footnote{\url{https://huggingface.co/castorini/rank_zephyr_7b_v1_full}} released by the author and run it ourselves to ensure the exact same settings. We also report the effectiveness of RankGPT, a zero-shot listwise reranked where GPT-4 is used as backbone. Due to budget constraints, we were unable to execute the experiments with RankGPT ourselves: we could only report the results of RankGPT obtained by \citet{sun-etal-2023-chatgpt} for TREC DL19 and DL20, and by \citet{su2025bright} for BRIGHT. Note that for the BRIGHT dataset, RankGPT was applied to a different implementation of BM25 from the one we used: the one used by RankGPT has a higher average nDCG@10 than our BM25.

\paragraph{Backbone LLMs.} For Setwise and Rank-R1, we explore base LLMs using instruction-tuned Qwen2.5 series models~\cite{qwen2025qwen25technicalreport} ranging from 3 billion to 14 billion parameters. RankZephyr is based on the Zephyr 7B backbone~\cite{tunstall2023zephyr}; RankGPT is based on OpenAI's GPT-4 model~\cite{openai2024gpt4technicalreport}.

\begin{table}[]
\centering
\begin{tabular}{l|l|l|l}
\hline
\textbf{Model}      & \textbf{Training}   & \textbf{DL19} & \textbf{DL20} \\ \hline
BM25    & zeroshot                           & .506               & .480  \\\hline
RankZephyr-7B                 & GPT4-distil.                 & .739               & .706    \\ 
RankGPT & Zeroshot & .756 & .706\\
\hline
Setwise-3B           & Zeroshot          & .371 & .317                     \\ 
Setwise-3B               & SFT          & .734 & .672                      \\ 
Rank-R1-3B            & Zeroshot          & .605 & .538                     \\ 
Rank-R1-3B            & GRPO             & .713 & .668                     \\\hline
Setwise-7B               & Zeroshot          & .675 &.636                      \\ 
Setwise-7B               & SFT          & .738 & .692                      \\ 
Rank-R1-7B             & Zeroshot          & .712 & .662                     \\ 
Rank-R1-7B             & GRPO             & .727 & .685                     \\ \hline
Setwise-14B              & Zeroshot          & .677& .648                    \\ 
Setwise-14B             & SFT          &  .729 & .689                     \\ 
Rank-R1-14B           & Zeroshot          & .679 & .652                     \\ 
Rank-R1-14B           & GRPO             & .714 & .691                     \\ \hline

\end{tabular}
\caption{TREC DL19 and DL20 nDCG@10 results. SFT=supervised fine tuned. GRPO trained on only 18\% of 400k data used to train SFT.}
\label{results-msmarco}
\end{table}

\section{Results}
\subsection{In-domain effectiveness}
In Table~\ref{results-msmarco}, we present the effectiveness of Rank-R1 variants 
on the TREC-DL19 and DL20 passage ranking datasets. 

We start by comparing Setwise and Rank-R1 under the zero-shot setting. The results suggest that incorporating the reasoning process into the Setwise method improves zero-shot ranking effectiveness. The improvements are particularly large on the 3B size model.

Next we consider the effects of training with GRPO. With GRPO training, Rank-R1 effectiveness increases, indicating that reasoning and answer generation (i.e. ranking) are enhanced by RL training. This improvement makes Rank-R1 comparable to the Setwise SFT trained on the full dataset and brings it closer to the effectiveness of the state-of-the-art RankZephyr.

\begin{figure}[]
	\centering
	\includegraphics[width=0.5\textwidth]{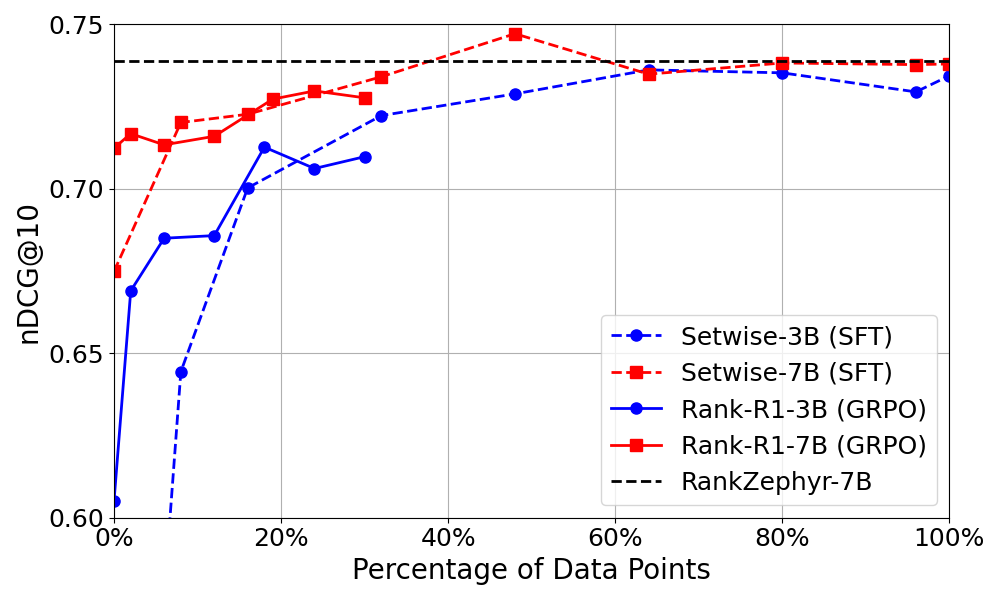}
	\caption{Data efficiency comparison between Setwise SFT and Rank-R1.}
	\label{fig:plots}
\end{figure}

\begin{table*}[]
\centering
\resizebox{1\textwidth}{!}{
    \begin{tabular}{l|l|l|l|l|l|l|l|l|l|l|l|l|l|l}
    \hline
    \textbf{Model}        & \textbf{Training}   & \textbf{Bio.} & \textbf{Earth.}& \textbf{Econ. }& \textbf{Psy.}& \textbf{Rob.}& \textbf{Stack.}& \textbf{Sus.}& \textbf{Pony}& \textbf{Leet.}& \textbf{AoPS}& \textbf{TheoT.}& \textbf{TheoQ.}& \textbf{Avg.} \\ \hline
    BM25   & zeroshot   &.182&.279&.164&.134&.109&.163&.161&.043&\textbf{.247}&.065&.021&.073&.137   
    \\
    \hline
    RankZephyr-7b     & GPT4-distill   & .219 & .237 & .144 & .103 & .076 & .137 & .166 & .065 & \textbf{.247} & .068 & .020 & .073 & .130                \\
    RankGPT4$^*$       & Zeroshot   &\textbf{.338}&.342&.167&\textbf{.270}&.223&\textbf{.277}&.111&\textbf{.156}&.034&.012&.086&.002&.170                \\\hline
    Setwise-3B     & Zeroshot & .143 & .175 & .120 & .102 & .077 & .079 & .154 & .053 & .154 & .017 & .042 & .021 & .095   \\ 
    
    Setwise-3B     & SFT & .220 & .188 & .104 & .115 & .091 & .058 & .167 & .057 & .099 & .040 & .034 & .038 & .101   \\  
    Rank-R1-3B     & Zeroshot & .137 & .173 & .119 & .152 & .100 & .066 & .178 & .037 & .077 & .040 & .060 & .025 & .097   \\  
    Rank-R1-3B    & GRPO    & .184 & .171 & .137 & .169 & .090 & .100 & .165 & .047 & .111 & .035 & .059 & .032 & .108   \\ \hline
    Setwise-7B    & Zeroshot& .236 & .223 & .161 & .171 & .149 & .092 & .183 & .063 & .149 & .041 & .104 & .056 & .136  \\  
     Setwise-7B     & SFT& .287 & .301 & .141 & .239 & .189 & .137 & .196 & .071 & .207 & .070 & .082 & .082 & .167   \\  
    Rank-R1-7B     & Zeroshot& .268 & .248 & .179 & .221 & .174 & .103 & .211 & .044 & .156 & .033 & .104 & .059 & .150   \\ 
    Rank-R1-7B     & GRPO    & .260 & .285 & .172 & .242 & .191 & .104 & .242 & .043 & .198 & .043 & .109 & .083 & .164 \\\hline
    Setwise-14B    & Zeroshot& .295 & .322 & .205 & .248 & .189 & .147 & .236 & .087 & .187 & .080 & .093 & .076 & .180   \\  
    Setwise-14B    & SFT& .220 & .293 & .154 & .230 & .201 & .157 & .203 & .062 & .194 & .095 & .099 & \textbf{.097} & .167\\
    Rank-R1-14B    & Zeroshot& .301 & .366 & \textbf{.221} & .246 & .217 & .154 & .250 & .090 & .170 & .091 & .116 & .092 & .193   \\ 
    Rank-R1-14B    & GRPO    &  .312 & \textbf{.385} & .212 & .264 & \textbf{.226} & .189 & \textbf{.275} & .092 & .202 & \textbf{.097} & \textbf{.119} & .092  & \textbf{.205}  \\  \hline
    \end{tabular}
}
\caption{BRIGHT nDCG@10 results. All methods rerank BM25 top-100 documents (First line). *: Results directly copied from the paper which uses a different BM25 ranking system (has a higher average nDCG@10 than our BM25).}
\label{tab:bright}
\end{table*}

\subsection{Effect of quantity of training data}

The results in Table~\ref{results-msmarco} for Rank-R1 trained with GRPO are obtained when using only 18\% of the MSMARCO training data (while SFT used all available training data). To explore whether longer training could further improve effectiveness, we continued training the 3B and 7B Rank-R1 models for an additional two days and evaluated checkpoints saved during training. We report the results in Figure~\ref{fig:plots}. In the figure, we also include results obtained when using SFT on incremental parts of the training data. 

From the figure, we observe that Rank-R1 requires significantly less data than Setwise SFT to achieve the same level of performance at early training stage -- however this data efficiency effect vanishes early on during the training phase. Passed 5-7\% of training data, in fact, the two training approaches tend to track each other. SFT has a clear advantage over GRPO in that it is by far less computationally expensive. On the other hand, GRPO adds new features to the reranker, introducing the ability to perform reasoning.

\subsection{Reasoning intensive out-of-domain effectiveness}
Next we consider results from our out-of-domain experiments, reported in Table~\ref{tab:bright}.

We observe that the SOTA RankZephyr reranker, which does not incorporate reasoning, does not provide better rankings than BM25 in most datasets from the BRIGHT benchmark -- effectively failing at the reranking task. This suggests that the BRIGHT benchmark poses a challenge for current SOTA LLM rerankers. 

On the other hand, Rank-R1 trained with GPRO outperforms or is on par with both zero-shot and Setwise SFT models in most cases. Notably, when using the 14B model, Setwise SFT effectiveness plateaued and even performed worse than its zero-shot counterparts, suggesting that the large model trained on the MSMARCO could not generalize to the BRIGHT using the standard Setwise approach. However, Rank-R1 based on the 14B model achieves the largest performance gain over Setwise SFT and even surpasses the GPT-4-based Listwise reranker baseline (RankGPT4). These results highlight that the reasoning process can help the model generalize to different domains and that strong reasoning abilities, along with larger model sizes, are crucial for LLM-based rerankers to be effective in reasoning-intensive ranking tasks.

\section{Analysis}

\subsection{Reward score v.s. Response length}
\begin{figure}[t]
	\centering
	\includegraphics[width=0.5\textwidth]{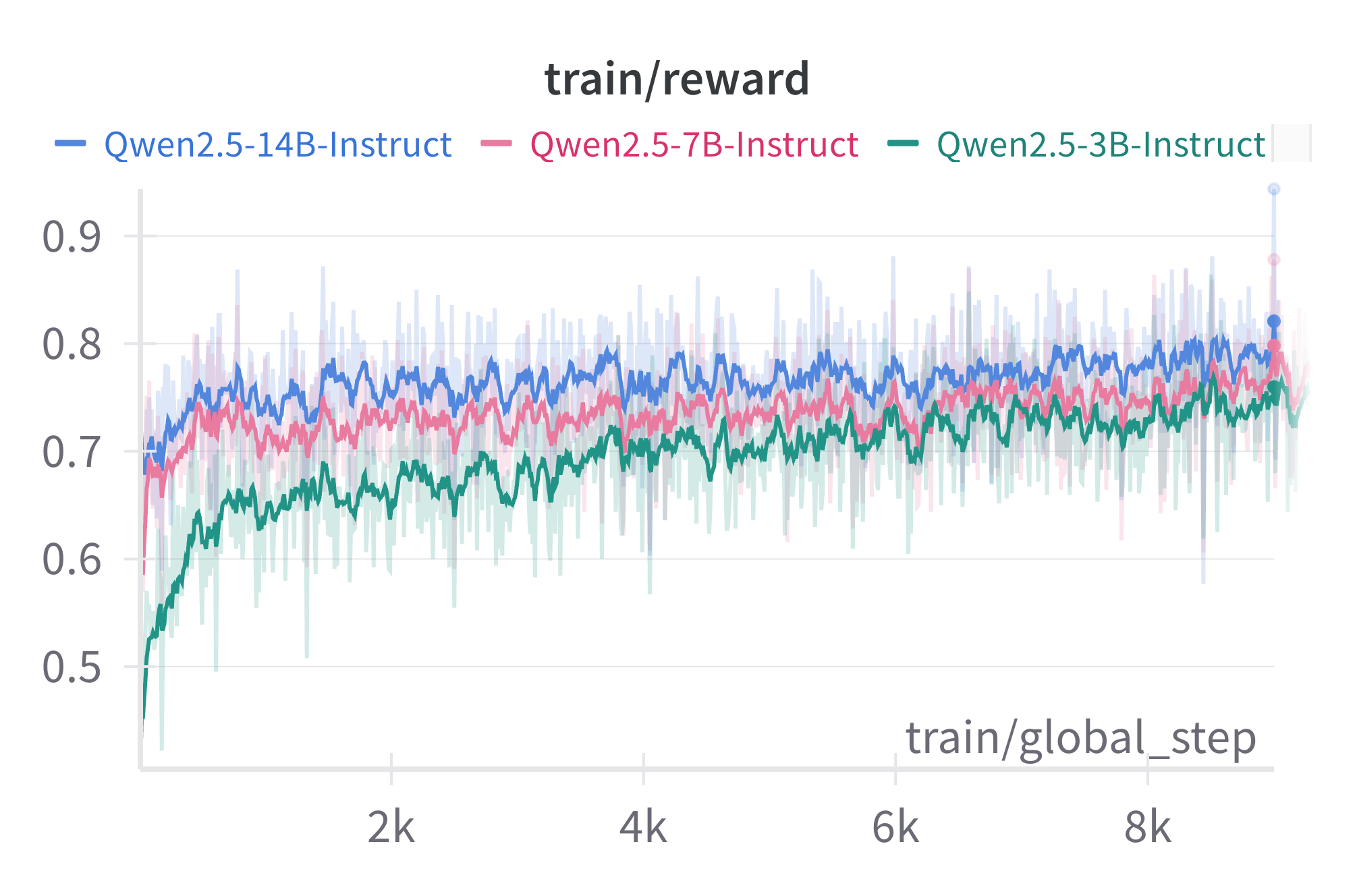}
	\includegraphics[width=0.5\textwidth]{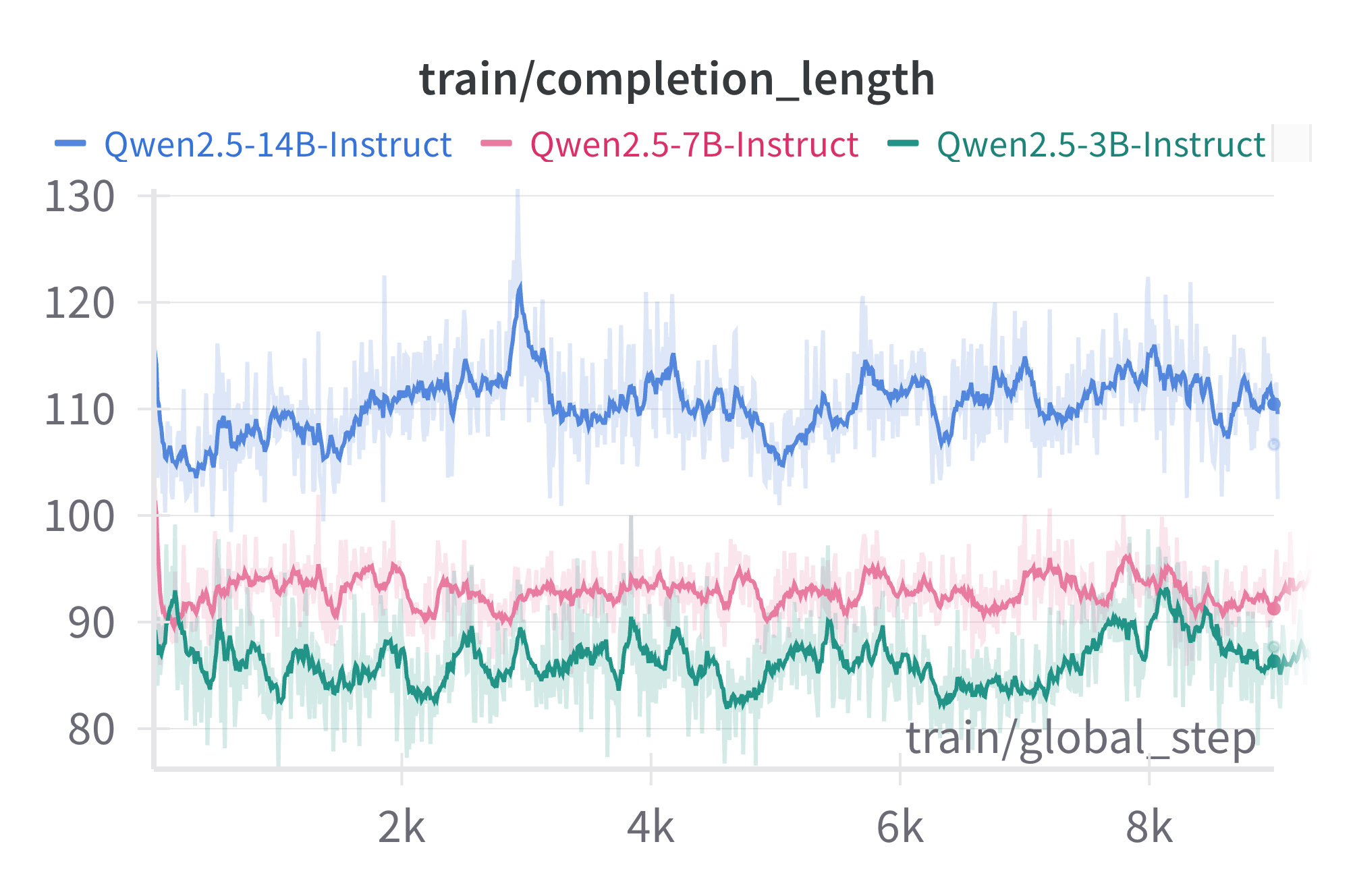}
	\caption{
		Rewards (top) and model completion length (bottom) obtained during GRPO training.}
	\label{fig:reward}
\end{figure}

In Figure~\ref{fig:reward}, we present the received reward values and model completion lengths logged during training for Rank-R1, across different model sizes. Rewards consistently increase throughout training, with smaller models showing a higher rate of increase, while larger models start with a higher initial reward.

Regarding completion length, larger models tend to generate longer responses; however, we do not observe a noticeable increase in length as training proceeds. This observation differs from the findings for DeepSeek-R1~\cite{deepseekr1}. This may be attributed to two factors. First, we initialize RL training from an instruction-tuned model rather than a base model, meaning the instruction model already follows a reasonable reasoning process. Second, the MSMARCO passage ranking dataset is relatively simple compared to tasks like math or coding, where a longer reasoning process is more essential. Thus, extensive reasoning may not be necessary for achieving high effectiveness in this task.

\subsection{Case study}
In Figure~\ref{fig:example}, we provide an example of Rank-R1's generation. We compare the outputs of the Zeroshot model and the model after GPRO training. Both models successfully follow the instruction by providing a reasoning process within the <think> span and predicting a relevant document label in the correct format. However, the Zeroshot model tends to merely describe what each document mentions and ultimately makes an incorrect prediction. In contrast, the GPRO-trained model focuses on the most relevant documents, compares them, and correctly selects the best one. In addition, we argue that Rank-R1's transparent reasoning process makes its predictions more explainable, which could be particularly important in domains such as medical document ranking.

\begin{figure}[t]
    \begin{tcolorbox}
    \small
    \textbf{Query:} common minerals list\\

    \textbf{Zeroshot Rank-R1 Completion:}\\
    <think>The query is "common minerals list". Document [3] talks about mica being a common rock-forming mineral. Document [19] mentions that chemical sedimentary rocks are formed through precipitation of minerals. However, it does not provide a comprehensive list of common minerals. Documents [1], [2], [4], [5], [6], [7], [8], [9], [10], [11], [12], [13], [14], [15], [16], [17], [18], and [20] do not mention anything related to common minerals.</think>\\
    <answer>[3]</answer>\\
    
    \textbf{GRPO-Trained Rank-R1 Completion:}\\ 
    <think>The query is "common minerals list". Document [3] talks about mica, which is a common rock-forming mineral. Document [4] also mentions common rock-forming minerals. Both documents seem relevant to the query. However, document [4] provides a list of common rock-forming minerals, which directly answers the query. </think>\\
    <answer>[4]</answer>
    
    \end{tcolorbox}
    \caption{Example of completion. The Rank-R1 used in this example is based on Qwen2.5-3B-Instruct. The GRPO-Trained Rank-R1 gives the correct answer.}
    \label{fig:example}
\end{figure}

\section{Conclusion}
In this paper, we introduce Rank-R1, which uses training based on the GPRO reinforcement learning strategy to enhance the reasoning process of an LLM-based Setwise reranker. Our training approach does not require human-annotated reasoning data; instead, it relies only on doument relevance assessments. We find that RL training performs similarly to supervised fine tuning on in-domain tasks.
However, in reasoning-intensive ranking tasks, Rank-R1 based on a 14B LLM achieves considerable higher effectiveness than the SOTA non-reasoning Listwise reranker, highlighting the importance of incorporating a reasoning process in document reranking. Moreover, this reasoning capability might improve the explainability of black-box LLM-based rerankers, and offer new affordances in terms of search engine result presentation and fruition.

We have made our code open-source at \url{https://github.com/ielab/llm-rankers/tree/main/Rank-R1}.



\bibliography{anthology,custom}

\appendix
\section{Prompt for Setwise reranker}\label{sec:appendix}
\begin{tcolorbox}
\small
SYSTEM:\\
A conversation between User and Assistant. The user asks a question, and the Assistant solves it. The assistant provides the user with the answer enclosed within <answer> </answer> tags, i.e., <answer> answer here </answer>.\\

USER:\\
Given the query: "\{query\}", which of the following documents is most relevant?\\
\textnormal{[1]} \{document1\}\\
\textnormal{[2]} \{document2\}\\
....\\
\textnormal{[20]} \{document20\}

Please provide only the label of the most relevant document to the query, enclosed in square brackets, within the answer tags. For example, if the third document is the most relevant, the answer should be: <answer>[3]</answer>.
\end{tcolorbox} 
We use the above prompt for both zero-shot and supervised fine-tuning of Setwise rerankers. The only difference from the prompt used for \textit{Rank-R1} is that the reasoning instructions are removed.

\section{Training hyper-parameters}\label{sec:appendix_train}
We use the TRL~\cite{vonwerra2022trl} library for both GPRP and SFT training. Both methods are trained with LoRA adapter~\cite{hu2022lora}. For SFT, we compute the cross-entropy loss only on the answer span and ignore the loss on the prompt tokens. The hyperparameters are listed in Table~\ref{parameters}, while other parameters follow the default settings of the TRL trainer.

\begin{table}[t]
\centering
\begin{tabular}{l|l|l}
\hline
\textbf{Parameter}      & \textbf{GRPO}   & \textbf{SFT}  \\ \hline
learning\_rate      & $1e-5$               & $1e-5$  \\
batch\_size      & 64               & 64  \\
optimizer &AdamW & AdamW \\
lora\_rank      & 16               & 16  \\
max\_prompt\_length & 4096               & 4096 \\
max\_completion\_length & 2048               & na \\
group\_size & 8               & na \\
\hline

\end{tabular}
\caption{Hyper-parameters for training.}
\label{parameters}
\end{table}



\end{document}